\documentclass[12pt,letterpaper]{article}
\pdfoutput=1
\usepackage[colorlinks=true,
linkcolor=red, 
citecolor=red,
filecolor=red,
urlcolor=red,
linktoc=page, 
pdfstartview=FitV,
bookmarksopen=true]{hyperref}
\usepackage[left=2cm,top=1cm,right=3cm,nohead]{geometry}
\usepackage[dvipsnames]{xcolor}
\usepackage{colortbl}
\usepackage[utf8]{inputenc} 
\usepackage{color}
\usepackage{epsfig,latexsym,amsfonts,mathtools,amsthm,amssymb,amsbsy,multirow,slashed,
	mathrsfs,wasysym,textcomp,subfigure,wrapfig,comment,bbold,array,longtable,multirow,setspace,amsmath,pifont}
\usepackage{tabularx}
\usepackage{graphicx}
\usepackage{setspace}
\usepackage{subfigure}
\usepackage{multirow}
\usepackage[bf]{caption}
\usepackage{braket}
\usepackage{comment}
\usepackage{float}
\usepackage{tikz}
\usetikzlibrary{mindmap,trees,shadows}
\colorlet{color1}{gray!25}
\usetikzlibrary{positioning}
\usetikzlibrary{intersections} 

\newlength{\PicScale}
\usepackage[UKenglish]{babel}
\usepackage[toc,page]{appendix}
\usepackage{hhline}
\usepackage{geometry}
\usepackage{cite}
\usepackage{datetime}
\usepackage{bbm} 
\usepackage{bm} 
\hypersetup{
	pdftitle={},
	pdfauthor={},
	pdfsubject={}
} 

\definecolor{Gray}{gray}{0.94}

\newcommand{\cmark}{\textcolor{Blue}{\ding{51}}}%
\newcommand{\xmark}{\textcolor{Red}{\ding{55}}}%

\newcolumntype{M}[1]{>{\centering\arraybackslash}m{#1}}
\newcolumntype{N}{@{}m{0pt}@{}}

\numberwithin{equation}{section}

\makeatletter
\def\@cline#1-#2\@nil{
	\omit
	\@multicnt#1
	\advance\@multispan\m@ne
	\ifnum\@multicnt=\@ne\@firstofone{&\omit}\fi
	\@multicnt#2
	\advance\@multicnt-#1
	\advance\@multispan\@ne
	\leaders\hrule\@height\arrayrulewidth\hfill
	\cr
	\noalign{\nobreak\vskip-\arrayrulewidth}}
\makeatother

\graphicspath{{./Images/}}

\begin{document}
\pagestyle{empty}
\begin{center}        
  {\bf\LARGE New Supersymmetric String Moduli Spaces from Frozen Singularities\\ [3mm]}

\large{ H\'ector Parra De Freitas
 \\[2mm]}
{\small  Institut de Physique Th\'eorique, Universit\'e Paris Saclay, CEA, CNRS\\ [-1mm]}
{\small\it  Orme des Merisiers, 91191 Gif-sur-Yvette CEDEX, France.\\[0.2cm] }

{\small \verb"hector.parradefreitas@ipht.fr"\\[-3mm]}
\vspace{0.3in}

\small{\bf Abstract} \\[3mm]\end{center}
The current classification of $\mathcal{N} = 1$ string theories in eight and seven dimensions is completely captured by K3 surfaces with F-Theory or M-Theory frozen singularities. In this note we show that there are inequivalent ways of freezing certain collections of singularities which have the same ADE type, and so there are more connected components in the moduli space than previously thought; namely, one more in eight dimensions and three more in seven dimensions.  We argue that the new component in eight dimensions decompactifies to a string theory in nine dimensions with rank 1 gauge group, which has been so far unidentified. Constructing and studying the stringy descriptions corresponding to these moduli spaces is the subject of a companion paper. 

\newpage



\setcounter{page}{1}
\pagestyle{plain}
\renewcommand{\thefootnote}{\arabic{footnote}}
\setcounter{footnote}{0}

\newpage
\section{Introduction}

The moduli space of string vacua with sixteen supercharges has a rich mathematical structure which is constrained enough by supersymmetry so as to allow detailed and exhaustive analyses of it, specially for large number of spacetime dimensions \cite{deBoer:2001wca}. Being very well studied for a long time, it serves as a simple testing ground for Swampland conjectures, a topic of current interest \cite{Montero:2020icj,Hamada:2021bbz,Bedroya:2021fbu,Cvetic:2020kuw}. Conversely, said interest has also motivated many recent refinements in our understanding of string theory in this regime, mainly with respect to gauge symmetries \cite{Cvetic:2021sjm,Cvetic:2022uuu,Lee:2021usk,Collazuol:2022jiy,Font:2020rsk,Font:2021uyw,Fraiman:2021soq,Fraiman:2021hma}. 

An important problem in this program is that of \textit{rank reduction}, first studied in \cite{Bianchi:1991eu}. In general, for $d$ spacetime dimensions, there is a connected component of the moduli space where the vacua have gauge groups of rank $26-d$. Depending on the duality frame we employ, there are different supersymmetry preserving ways of transforming these into other vacua whose gauge groups have lower ranks, and hence belong to other connected components of the moduli space. A few examples are:
\begin{enumerate}
	\item The $E_8\times E_8 \rtimes \mathbb{Z}_2$ heterotic string on $S^1$ can be orbifolded by an exchange of the $E_8$ factors together with a half-shift along the circle to produce a theory, known as the CHL string \cite{Chaudhuri:1995dj, Chaudhuri:1995bf}, with gauge group rank reduced by 8. This procedure can be generalized in lower dimensions to various other rank reductions. In eight dimensions, the CHL string is T-dual to the $Spin(32)/\mathbb{Z}_2$ heterotic string on a $T^2$ without vector structure\cite{Witten:1997bs}. 
	
	\item In Type I' string theory \cite{Polchinski:1995df} on $S^1$ one may exchange one of the two $O8^-$-planes with all 16 $D8$-brane pairs on top by an $O8^+$-plane, reducing the rank of the gauge group by 16. A type II orientifold on $T^2$ with four $O7^-$ planes is more versatile and exchanges of this type allow to reduce the rank of the gauge group by 8 or 16.
	
	\item M-theory can be compactified on a cylinder, a Möbius strip or a Klein bottle, realizing nine dimensional theories with gauge groups of rank 17, 9 or 1, respectively \cite{Dabholkar:1996pc}.
	
	\item In eight dimensions, F-theory on an elliptic K3 surface with section can have one or two partially frozen singular fibers of ADE type $D_{8+n}$, respectively reducing the gauge group rank by 8 or 16 \cite{Bhardwaj:2018jgp}. Similarly, in seven dimensions, M-theory on a K3 surface can have partially frozen ADE singularities of type $D_{4+n}$ or $E_{6+n}$ which reduce the gauge group rank by 8, 12, 14, 16 or 18 \cite{deBoer:2001wca}. 
\end{enumerate}
All known theories with rank reduction by 8 are dual. In general, however, there can be inequivalent theories with the same rank reduction. An example of this is already available in nine dimensions, where Type I' on $S^1$ with $O8^\pm$-planes is not equivalent to M-theory on a Klein bottle \cite{Aharony:2007du}. However, when they are compactified to eight dimensions, they form part of the same moduli space. As a result, we are left with only one moduli space component with gauge group rank 2 in eight dimensions; no alternative constructions allowed by having two compact dimensions are known.

In this note we wish to clarify that the moduli space of F-Theory on K3 with two frozen singularities has in fact two disconnected components. As opposed to $D_8$, the lattice $D_8\oplus D_8$ can be embedded into $\Gamma_{2,18}$ in two inequivalent ways, one related to the gauge group $Spin(16)^2/\mathbb{Z}_2$ and the other to $Spin(16)^2$. The former is dual to a circle compactification of the two rank 1 theories mentioned above, while the later is not.

Along the same lines, we identify different moduli space components of M-theory on K3 corresponding to inequivalent configurations of frozen ADE singularities with the same ADE type. Of these, one was already conjectured to exist in \cite{deBoer:2001wca} for frozen $4\, D_4$, since there are two inequivalent configurations of $O6^\pm$-planes in a seven dimensional type II orientifold with gauge group rank 3. Our results suggest that this theory is physically equivalent to the compactification of the alternative component in eight dimensions discussed above. The remaining alternative configurations are for $3 \, E_6$ and $D_4 + 2 E_7$, of which we find one in each case. 

\begin{table}
	\centering
	\def\arraystretch{1.4}
	\resizebox{360pt}{!}{%
		\begin{tabular}{|c||c|c|c|}\hline
			rank reduction& $d=9$ &  $d=8$ & $d=7$\\ \hline 
			$8$&$E_8$&$Spin(16)$&$Spin(8)^2$\\ \hline
			\multirow{3}{*}{$16$}&$E_8^2$&\multirow{2}{*}{$\frac{Spin(16)^2}{\mathbb{Z}_2}$}& \multirow{2}{*}{$\frac{Spin(8)^4}{\mathbb{Z}_2\times \mathbb{Z}_2}$}\\ \cline{2-2}
			&$\frac{Spin(32)}{\mathbb{Z}_2}$&&\\\cline{2-4}
			&$Spin(32)$~\ding{75}&$Spin(16)^2$~\ding{75}&$\frac{Spin(8)^4}{\mathbb{Z}_2}$~\ding{75}\\\hline
			$12$&-&-&$E_6^2$\\\hline
			$14$&-&-&$E_7^2$\\\hline
			\multirow{2}{*}{$16$} &-&-& $E_8^2$\\\cline{2-4}
			&-&-&$E_8^2{}'$\\\hline
			\multirow{5}{*}{$18$}&-&-&$\frac{E_6^3}{\mathbb{Z}_3}$\\\cline{2-4}
			&-&-&$E_6^3$~\ding{75}\\\cline{2-4}
			&-&-&$\frac{Spin(8)\times E_7^2}{\mathbb{Z}_2}$\\\cline{2-4}
			&-&-&$Spin(8)\times E_7^2$~\ding{75}\\\cline{2-4}
			&-&-&$Spin(8)\times E_6 \times E_8$\\\hline
		\end{tabular}
	}
	\caption{Gauge groups corresponding to the singularity freezings producing the different moduli space components in $d \geq 7$ string theories with 16 supercharges. The leftover rank in each case is equal to $26-d-$rank reduction. All fundamental groups are diagonal. The two $E_8^2$ cases differ at the level of the M-Theory 3-form field background which produces the freezing. Items labeled with \ding{75} correspond to new components. The case of seven dimensions refines Table 1 of \cite{deBoer:2001wca}.} \label{tab:frozen}
\end{table}
Moreover, the lattices associated to frozen singularities defining a moduli space component and its torus compactifications obey clear patterns which suggest that the configuration corresponding to $Spin(16)^2$ uplifts to an as yet unindentified rank 1 moduli space in nine dimensions. To wit, $Spin(16)^2/\mathbb{Z}_2$ uplifts to $Spin(32)/\mathbb{Z}_2$, while $Spin(16)^2$ uplifts to $Spin(32)$. Both of these gauge groups exist in nine dimensions, and so we are led to conjecture that there are in fact three and not two rank 1 nine dimensional string theories, perhaps described best in terms of a real elliptic K3 surface \cite{Cachazo:2000ey} with frozen singularities. In this picture, the CHL string corresponds to $E_8$, M-Theory on a Klein bottle to $2\, E_8$, and Type I' with $O8^\pm$-planes and the new theory to inequivalent $D_{16}$'s. We believe that with this refinement all the components of the moduli space of $d \geq 7$ string theories and 16 supercharges are captured by the frozen singularity picture, which we list in Table \ref{tab:frozen}. 

Stringy descriptions of the new components predicted here turn out to exist and are given e.g. by turning on discrete theta angles in known string backgrounds; their construction and analysis is the subject of the companion paper \cite{MP}.

The paper is organized as follows: In Section \ref{sec:2} we set up F-Theory with frozen singularities and explain why there are two rank 2 components. Then we examine the gauge symmetry enhancements therein and propose charge lattices for the theories. In Section \ref{sec:3} we repeat this analysis for M-Theory on K3 with frozen singularities. In Section \ref{sec:4} we discuss how the two rank 2 components in eight dimensions may be uplifted to nine dimensional theories. In Appendix \ref{app:lats} we consider the problem of uniqueness for each of the components in Table \ref{tab:frozen}.  
\section{F-Theory with frozen singularities (8d)}
\label{sec:2}
The data characterizing an F-theory vacuum in eight dimensions is encoded in the Weierstrass model for an elliptically fibered K3 surface with section, 
\begin{equation}\label{Weierstrass}
	y^2 = x^3 + f(u,v)xz^4 + g(u,v)z^6\,,
\end{equation}
where $x,y,z$ are the homogeneous coordinates on the fiber ambient space $\mathbb{P}^{231}$, $u,v$ are the homogeneous coordinates on the base $\mathbb{P}^1$, and $f, g$ are arbitrary polynomials of degree 8 and 12. There are generically 24 singular fibers located at the points of the base given by the zeros of the discriminant
\begin{equation}
	\Delta(u,v) = 4f^3(u,v) + 27 g^2(u,v).
\end{equation}
Varying the complex structure moduli of the K3, the zeros of $\Delta$ may degenerate, and the singularities at such points worsen. Depending on the monodromies around these singularities, the gauge algebra of the physical theory can get enhanced from the generic $18\,\mathfrak{u}(1)$ to various non-abelian algebras. 

\subsection{One frozen singularity}

The possible singularities that can occur were classified by Kodaira and Néron and are well known by now. Of particular importance to us is that of type $I_{4+n}^*$, which gives rise to an $\mathfrak{so}(16+2n)$ gauge algebra. It has an alternative variant $\hat I_{4+n}^*$ in F-Theory whose gauge algebra is instead $\mathfrak{sp}(n)$ \cite{Witten:1997bs}, but is indistinguishable from $I_{4+n}^*$ at the level of the Weierstrass model \eqref{Weierstrass}. For $n = 0$, it does not produce gauge symmetry, and cannot be split into other singularities; the 8 complex structure moduli whose variation would produce this effect are frozen. A nice explanation of these facts from the type II orientifold point of view can be found in \cite{Bhardwaj:2018jgp}. 

On the other hand, gauge symmetry enhancements are known to correspond to primitive sublattices of the even self-dual Lorentzian lattice $\Gamma_{2,18}$. In particular, the moduli space of F-theory with generic gauge algebra $\mathfrak{so}(16)$ corresponds to a primitive embedding of the lattice $D_8$ into $\Gamma_{2,18}$, which is unique up to automorphisms (see Appendix \ref{app:lats}). Therefore, all configurations with a generic $I_{4}^*$, possibly inside an $I_{4+n}^*$, can be deformed into each other without breaking the associated $\mathfrak{so}(16)$ subalgebra, and the moduli space of vacua with one partially frozen $I_{4+n}^*$ is unique. This corresponds to a type II orientifold on $T^2$ with one $O7^+$-plane and three $O7^-$-planes, and is dual to the CHL string in eight dimensions.

\subsection{Two frozen singularities}
\label{ss:2f}

It is also possible to have two singular fibers $I_{4+n}^*$ and $I_{4+m}^*$ at the same time, which can be partially frozen. In this case, however, not all such freezings are equivalent. To see this, note that a generic setup $2 \, I_4^*$ has gauga algebra $2\, \mathfrak{so}(16)$, and so it corresponds to a primitive sublattice of $\Gamma_{2,18}$ whose root sublattice is $2\, D_8$. It turns out that there are two such lattices, 
\begin{equation}\label{lambdas}
\begin{split}
	\Lambda_1 &= W_{Spin(16)\times Spin(16)/\mathbb{Z}_2}\,,\\
	\Lambda_2 &= W_{Spin(16)\times Spin(16)}\,.
\end{split}	
\end{equation}
$\Lambda_1$ is the weight lattice of $Spin(16)\times Spin(16)/\mathbb{Z}_2$, where the fundamental group is generated by the diagonal spinor class $k = (s,s)$ in the center of $Spin(16)\times Spin(16)$, and $\Lambda_2$ is the weight lattice of the simply connected $Spin(16)\times Spin(16)$, or equivalently its root lattice. There are therefore two distinct moduli spaces with generic gauge algebra $2\, \mathfrak{so}(16)$: one with generic gauge group $Spin(16)\times Spin(16)/\mathbb{Z}_2$ and the other with $Spin(16)\times Spin(16)$. At the level of the elliptic K3 geometry, the first generically has an order two section in its Mordell-Weil group, while the other does not. 

To see that one of these configurations corresponds to a perturbative type II orientifold on $T^2$, consider the dual $Spin(32)/\mathbb{Z}_2$ heterotic string on $T^2$ with Wilson lines
\begin{equation}
	A_1 = (0^8,\tfrac12^8)\,, ~~~~~ A_2 = (0^{16})\,.
\end{equation}
The first Wilson line breaks the gauge group to $Spin(16)^2/\mathbb{Z}_2$. In particular, we have that
\begin{equation}
	A_1 \cdot (\tfrac12^{16}) = 2 \in \mathbb{Z}\,,
\end{equation}
and so the massive spinor in the weight lattice is preserved. We can then read the values $(A_1^a, A_2^a)$, with $a = 1,...,16$, as the coordinates of the sixteen $D7$-brane pairs in the dual orientifold $T^2$. Eight pairs are located at one $O7^-$-plane and the eight pairs at another, while two $O7^-$-planes remain naked. Exchanging the two $O7^- + 8 D8$ stacks by $O7^+$ planes realizes the rank 2 theory orientifold theory, referred to as $(+,+,-,-)$ in \cite{deBoer:2001wca}.

To get instead $Spin(16)\times Spin(16)$ we must set the second Wilson line to
\begin{equation}
	A_2 = (1,0^{15})\,,
\end{equation}  
which preserves the gauge algebra but breaks the massive spinor in $\Lambda_2$, since $A_2 \cdot (\tfrac12^{16}) = \tfrac12 \notin \mathbb{Z}$. This value for $A_2$ is however beyond the region in moduli space which can be perturbatively described by a Type II orientifold. 

\subsection{Effect on the gauge groups}

The gauge symmetry groups that can occur in F-Theory with one frozen singularity have been studied exhaustively from the point of view of the dual CHL string in \cite{Font:2021uyw,Cvetic:2021sjm}, and more recently from the point of view of type IIB string junctions \cite{Cvetic:2022uuu}. A map relating all rank 18 groups with $Spin(16)$ subgroup with all rank 10 groups, taking into account their topology, was also obtained in \cite{Cvetic:2021sjm}. In the F-Theory picture, this map simply transforms $I_{4+n}^*$ into $\hat I_{4+n}^*$, so that an $\mathfrak{so}(16+2n)$ algebra transforms into an $\mathfrak{sp}(n)$. The fundamental group $H$ of the full gauge group transforms at the level of its generators, but the result $H'$ is isomorphic to $H$ in every case, so that for example the gauge group $Spin(32)/\mathbb{Z}_2$ transforms into $Sp(8)/\mathbb{Z}_2$. Therefore, the fundamental group is still given by the torsion of the Mordell-Weil group of sections MW as for the standard component \cite{Aspinwall:1998xj,Mayrhofer:2014opa}.

\subsubsection{Component $\mathcal{M}_{2D_8}$}
\label{sss:M81}
Now we wish to extend this picture to the moduli space components with gauge group rank 2, starting with the one with frozen singularities associated to the gauge group $Spin(16)^2/\mathbb{Z}_2$ which we denote by $\mathcal{M}_{2D_8}$. The only gauge groups that arise in the fibrations that can be mapped to this moduli space are the following \cite{Shimada2000OnEK,Font:2020rsk}:
\begin{eqnarray}
	\resizebox{360pt}{!}{%
		\def\arraystretch{1.5}
		\begin{tabular}{|c||c|c|c|}\hline
		$\tilde G$ & $Spin(16)^2\times U(1)^2$ & $Spin(16)^2\times SU(2)\times U(1)$ & $Spin(16)^2\times SU(2)^2$\\ \hline
		$H$&$\mathbb{Z}_2$ & $\mathbb{Z}_2$ & $\mathbb{Z}_2\times \mathbb{Z}_2$ \\ \hline
		$\{k_i\}$& $(s,s)$ & $(s,s,0)$ & $(s,s,0,0)\,, (c,v,1,1)$ \\ \hline 
		\end{tabular}
	}\,
\end{eqnarray}
where $\tilde G$ is the universal cover, $H$ the fundamental group and $\{k_i\}$ the generators of $H$, whose entries correspond to the non-abelian factors in $\tilde G$. At the level of the algebras, we have that $2\,\mathfrak{so}(16) \mapsto \emptyset$ in each case, corresponding to the freezing of the fibers $2\, I_4^* \to 2\, \hat I_4^*$. In contrast with the case of one frozen singularity, however, we see that $H$ must transform. The generic gauge group in the rank 2 moduli space should be $U(1)^2$ and so $H' = \emptyset$. Indeed we define $H$ as the fundamental group of the \textit{non-abelian} part of the gauge group, ignoring the topological aspects of the abelian part \cite{Cvetic:2021sjm}. There is still an order two section in the elliptic K3, but it does not intersect the singular fibers that result in gauge symmetries. This can be seen clearly in the two enhancements in the table above. The first enhancement is therefore just $SU(2)$, but the maximal one is more interesting. We expect it to be $SU(2)^2/\mathbb{Z}_2$, with $\mathbb{Z}_2$ diagonal, or equivalently $SO(4)$. An explicit computation in \cite{Cvetic:2022uuu} confirms this expectation.

Now, two constraints for non-simply-connected gauge groups $G$ in quantum gravity theories were presented recently in \cite{Montero:2020icj} and \cite{Cvetic:2020kuw}, respectively, and it is instructive to see that the results here satisfy them. The former requires that in the case $G$ only has real representations, it must satisfy $\text{dim}(G) + \text{rank}(G) = 0 \mod 8$. The group $SO(4)$ has dimension 28 and rank 4 and so passes the test. The later requires in particular for $G$ with universal cover $SU(n_1)\times\cdots \times SU(n_s)$ and cyclic $\pi_1(G)$ that
\begin{equation}\label{consCv}
	\sum_{a = 1}^s \frac{n_a-1}{2n_a} k_a^2 m_a \in \mathbb{Z}\,,
\end{equation} 
where $m_a$ is the level of the $a$-th factor in the associated current algebra and $k_a$ is the $a$-th component of the generator of $\pi_1(G)$. For $SU(2)^2/\mathbb{Z}_2$, \eqref{consCv} reduces to
\begin{equation}\label{consCv'}
	\frac{m_1+m_2}{4} \in \mathbb{Z}\,.
\end{equation}
Given that the two $SU(2)$ factors are in the same footing, we take $m_1 = m_2 = m$. It is clear then that $m \geq 2$. On the other hand we have that the gauge group contributes to the left-moving central charge as
\begin{equation}
    c_G = 2 \times \frac{m ~ \text{dim}G}{m+h^\vee}\,.
\end{equation}
For $m = 2$ this gives $c_G = 3$, which together with the contribution $c = 6+3$ coming from the transverse bosons and their worldsheet superpartners saturates the bound $c_L \leq 12$ for the Type II string. 
\subsubsection{Component $\mathcal{M}_{2D_8}'$}

Let us now consider the ``new" moduli space component $\mathcal{M}_{2D_8}'$, with frozen singularities associated to the gauge group $Spin(16)^2$. Prior to freezing we have only the following enhancements: 
\begin{eqnarray}
\resizebox{360pt}{!}{%
	\def\arraystretch{1.5}
	\begin{tabular}{|c||c|c|c|}\hline
	$G$ & $Spin(16)^2\times U(1)^2$ & $Spin(16) \times Spin(18)\times U(1)$ & $Spin(18)^2$\\ \hline
	\end{tabular}
}\,
\end{eqnarray}
Since there is no torsional MW at all, the gauge groups that result after freezing the singularities are $U(1)^2$, $SU(2)\times U(1)$ and $SU(2)^2$. These results are also in agreement with \cite{Cvetic:2022uuu}. 

At the level of gauge algebras, we see that the symmetry enhancements in this moduli space are exactly the same as those of $\mathcal{M}_{2D_8}$. The difference is only seen in the gauge group topology, which involves massive states. 

\subsection{Charge lattices}
\label{ssec:lattices8}
The charge lattice associated to the moduli space of F-theory with one frozen singularity is the orthogonal complement of the lattice $D_8$ in $\Gamma_{2,18}$, namely $\Gamma_{2,2}\oplus D_8$. In fact, all the charge lattices constructed in \cite{deBoer:2001wca} for theories with 16 supercharges down to six dimensions are orthogonal complements of the sublattice of the Narain lattice corresponding to the frozen moduli, as first observed by Mikhailov \cite{Mikhailov:1998si}. It is therefore natural to guess that for the two moduli spaces with gauge group rank 2 in eight dimensions the charge lattices are respectively the orthogonal complements of $\Lambda_1$ and $\Lambda_2$ (cf. Eq. \eqref{lambdas}) in $\Gamma_{2,18}$,  which take the form
\begin{eqnarray}
	\Gamma_{2D8} &=& \Gamma_{1,1}\oplus \Gamma_{1,1}(2)\,,\label{gamma1}\\
	\Gamma_{2D_8}' &=& \Gamma_{2,2}(2)\,,\label{gamma2}
\end{eqnarray}
where the notation $(n)$ means that the lattice vectors are dilated by a factor of $\sqrt{n}$. This $\Gamma_{2D_8}$ indeed matches the proposed charge lattice in \cite{Cvetic:2022uuu}, and in fact can be obtained by shaving off a $\Gamma_{1,1}(2)$ from the charge lattice of F-Theory on $(T^4 \times S^1)/\mathbb{Z}_2$ \cite{deBoer:2001wca}, hence it corresponds to a decompactification limit thereof. Here we emphasize however that there are two distinct moduli space components in question including one with a different charge lattice $\Gamma_{2D_8}'$.

Let us now look at how the root lattices of the maximal symmetry enhancements are realized in each case. For $\Gamma_{2D_8}$ we note first that it corresponds to a compactification of the nine dimensional theory with charge lattice $\Gamma_{1,1}$ (for example the AOA string) which adds the lattice $\Gamma_{1,1}(2)$. One enhancement to $\mathfrak{su}(2)\oplus \mathfrak{u}(1)$ can therefore be associated to the $\mathfrak{su}(2)$ of the nine dimensional theory given by massless states with charge vectors
\begin{equation}
	\ket{\alpha_1} = \ket{n_1,n_2,w^1,w^2} = \ket{1,0,1,0}\,,
\end{equation}
where the lattice inner product is given by $n_1 w'^1 + n_1' w^1 + n_2 w'^2 + n_2' w^2$. Now we must take e.g. $n_2 \in \mathbb{Z}$ and $w^2 \in 2\mathbb{Z}$ corresponding to $\Gamma_{1,1}(2)$. Recall that in the CHL string when we compactify the theory from nine to eight dimensions, the charge lattice $\Gamma_{1,9}$ is extended to $\Gamma_{1,9} \oplus \Gamma_{1,1}(2)$, and the added factor hosts a vector $\ket{1,2}$ with squared norm 4 giving an $\mathfrak{su}(2)$ at level 1 which is distinguished from those enhancements obtained from the rest of the lattice, and is in fact responsible for the appearance of symplectic gauge groups. In the present case however the enhancement to $2\,\mathfrak{su}(2)$ is symmetric in both factors so that states with charge vectors $\ket{0,1,0,2}$ in $\Gamma_{2D_8}$, which cannot be mapped to $\ket{1,0,1,0}$, are expected to be always massive. In other words, by tuning the moduli of the compactification circle alone one cannot get a symmetry enhancement. We take the second $\mathfrak{su}(2)$ to have instead the charge vectors
\begin{equation}
	\ket{\alpha_2} = \ket{1,1,-1,2}\,.
\end{equation} 

With the choices for $\alpha_1$ and $\alpha_2$ for the root lattice $L$ of $2\, \mathfrak{su}(2)$ let us now verify that the gauge group is indeed $SU(2)^2/\mathbb{Z}_2 \simeq SO(4)$, which amounts to proving that all the states in $\Gamma_{2D_8}$ sit either in the adjoint or the fundamental representation of both $SU(2)$'s simultaneously. To this end, first consider the orthogonal complement $L^\perp$ of $L$ in $\Gamma_{2D_8}$, generated by the vectors
\begin{equation}
	\ket{\beta_1} = \ket{1,0,-1,2}\,, ~~~~~ \ket{\beta_2} = \ket{0,1,0,-2}\,.
\end{equation}
The lattice $L\oplus L^\perp$ has determinant $4\times 4 = 16$, and so the vectors that extend it to $\Gamma_{2D_8}$, which has determinant $4$, correspond to a vector $\ket{v}$ such that $2\ket{v} \in L \oplus L^\perp$ together with all translations of $\ket{v}$ by elements of $L \oplus L^\perp$. We may take
\begin{equation}
	\ket{v} = \ket{0,0,1,0}\,,
\end{equation}
which sits in the fundamental representation of both $SU(2)$'s as expected. The remaining extra vectors also satisfy this property and so $Spin(4)$ spinors are not present and the gauge group is $SO(4)$. 

The case of $\Gamma_{2D_8}'$ is simpler. The most natural choice is to take each $SU(2)$ to correspond to each $\Gamma_{1,1}(2)$ factor. In this way, each $SU(2)$ is simply connected, as expected, since each $SU(2)$ root lattice can be completed to $\Gamma_{1,1}(2)$ by adding the vector corresponding the fundamental representation. In Section \ref{sec:4} we will argue that this lattice can be lifted to the charge lattice $\Gamma_{1,1}(2)$ corresponding to a new nine dimensional moduli space with gauge group rank 1 apart from the two studied in \cite{Aharony:2007du}.

\section{M-Theory with frozen singularities (7d)}
\label{sec:3}
Now we wish to extend the discussion of inequivalent theories with the same type of frozen singularities to seven dimensions, where the relevant setting is M-Theory on a K3 surface. The standard component, with gauge groups of rank 19, is just the moduli space of K3 surface metrics. At special points, this surface develops du Val (or ADE) singularities, whose neighborhoods are of the form $\mathbb{C}^2/\Gamma_\mathfrak{g}$, with $\Gamma_{\mathfrak{g}}$ a subgroup of $SU(2)$ related to an ADE algebra $\mathfrak{g}$. The presence of such singularities enhance certain $\mathfrak{u}(1)$'s of the gauge algebra of the theory to a direct sum of their associated ADE algebras.

A partially frozen singularity in this case is an ADE singularity on which the 3-form field $C_3$ is turned on such that
\begin{equation}
	\int_{S^3/\Gamma_\mathfrak{g}} C_3 = \frac{p}{q} \mod 1 \neq 0\,,
\end{equation}
with $S^3/\Gamma_\mathfrak{g}$ envolving it. Depending on the ADE type, certain values for $p$ and $q$ are allowed, and the gauge algebra that is realized is transformed into one of lower rank and possibly non-simply-laced. What concerns us here is that these partial freezings lead to different connected components in the moduli space and can be realized in subspaces of the standard component with generic ADE configurations of the following type:
\begin{equation}\label{fr1}
	2\, D_4^{1/2}\,, ~~~ E_6^{1/3}+E_6^{2/3}\,, ~~~ E_7^{1/4}+E_7^{3/4}\,, ~~~ E_8^{1/5}+E_8^{4/5}\,, ~~~ E_8^{1/6}+E_8^{5/6}\,,
\end{equation}
or
\begin{equation}\label{fr2}
	4 \, D_4^{1/2}\,, ~~~ 3\, E_6^{1/3}\,, ~~~ D_4^{1/2} + 2 \, E_7^{1/4}\,, ~~~ D_4^{1/2} + E_6^{1/3} + E_8^{1/6}\,, 
\end{equation}
where the superindex denotes the value of $p/q$ for each singularity. As for the eight dimensional case, these singularities are related to sublattices of $\Gamma_{3,19}$ whose root sublattices are of the corresponding ADE type. 

It can be shown that the root lattices for the singularities in \eqref{fr1} are embedded uniquely into $\Gamma_{3,19}$, and that there are no overlattices thereof (such as the weight lattice of ${E_6^2/\mathbb{Z}_3}$) in $\Gamma_{3,19}$, so that the associated moduli space components are unique (see Appendix \ref{app:lats}). In \eqref{fr2}, this statement holds for the last configuration but not for the others! Instead, we find them to have two components each. In the following we consider each case in detail. 

\subsection{Components with frozen $4\, D_4$}

As in Section \ref{ss:2f}, we can easily find different gauge groups with the same algebra $4\, \mathfrak{so}(8)$ by considering the dual heterotic string on $T^3$. In this case there are three, with the same values for two Wilson lines:
\begin{equation}\label{WL12}
\begin{split}
	A_1 &= (0^4\,, 0^4\,, \tfrac12^4\,,\tfrac12^4)\,,\\
	A_2 &= (0^4\,, \tfrac12^4\,, 0^4\,, \tfrac12^4 )\,,
\end{split}
\end{equation}
which break $Spin(32)/\mathbb{Z}_2$ to $Spin(8)^4/\mathbb{Z}_2^2$ with fundamental group
\begin{equation}
	\mathbb{Z}_2^2 = \{(0,0,0,0), (s,s,s,s), (v,v,v,v), (c,c,c,c) \}\,.
\end{equation}
The nontrivial elements in this group correspond respectively to the $\Gamma_{16}$ vectors
\begin{equation}
u_s = (\tfrac12^{16})\,, ~~~~~ u_v = (0^3,1,-1,0,0,1,-1,0,0,1,-1,0^3)\,, ~~~~~ u_c = (\tfrac12^{15},-\tfrac12)\,.
\end{equation}
Then there are different choices of $A_3$ which preserve the gauge algebra but may change the topology of the gauge group, namely
\begin{equation}\label{WL3}
	\begin{split}
	A_3^{(1)} = (0,0,0,0,0^{12}):~& ~~~~~ \mathbb{Z}_2^2 \mapsto \mathbb{Z}_2^2\,,\\
	A_3^{(2)} = (\tfrac12,\tfrac12,\tfrac12,\tfrac12, 0^{12}):~& ~~~~~ \mathbb{Z}_2^2 \mapsto \mathbb{Z}_2 = \{(0,0,0,0),(s,s,s,s)\}\,,\\
	A_3^{(3)} = (1,0,0,0,0^{12})~:~& ~~~~~ \mathbb{Z}_2^2 \mapsto \mathbb{Z}_2 = \{(0,0,0,0),(v,v,v,v)\}\,,\\
	A_3^{(4)} = (\tfrac12,\tfrac12,\tfrac12,-\tfrac12, 0^{12}):& ~~~~~ \mathbb{Z}_2^2 \mapsto \mathbb{Z}_2 = \{(0,0,0,0),(c,c,c,c)\}\,.
	\end{split}
\end{equation}
The weight lattices corresponding to the last three configurations are all isomorphic due to the triality property of $D_4$, but they seem to be embedded into $\Gamma_{3,19}$ in inequivalent ways; we find no T-dualities that relate the different configurations of Wilson lines. Regardless, they have isomorphic orthogonal complements and so the theories they define have the same charge lattice and symmetry enhancing patterns. We conclude therefore that even if there were inequivalent configurations of $4\, D_4$ singularities associated to $Spin(8)^4/\mathbb{Z}_2$, freezing them would produce physically equivalent theories. 

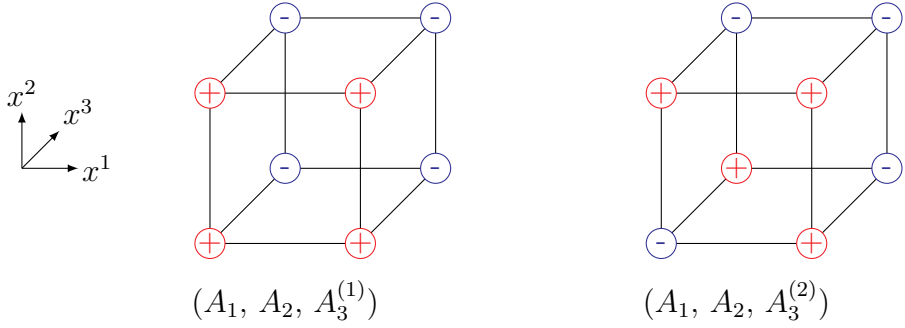
\begin{figure}
\centering
\begin{tikzpicture}
\draw(0,0)--(0,2)--(2,2)--(2,0)--(0,0);
\draw(1,1)--(1,3)--(3,3)--(3,1)--(1,1);
\draw(0,0)--(1,1);
\draw(0,2)--(1,3);
\draw(2,2)--(3,3);
\draw(2,0)--(3,1);
\draw[Red,fill=White](1,1)circle(0.2)node{+};
\draw[Red,fill=White](0,2)circle(0.2)node{+};
\draw[Red,fill=White](2,2)circle(0.2)node{+};
\draw[Red,fill=White](2,0)circle(0.2)node{+};
\draw[Blue,fill=White](0,0)circle(0.2)node{-};
\draw[Blue,fill=White](1,3)circle(0.2)node{-};
\draw[Blue,fill=White](3,3)circle(0.2)node{-};
\draw[Blue,fill=White](3,1)circle(0.2)node{-};
\draw(1,-0.75)node{$(A_1,\,A_2,\,A_3^{(2)})$};

\begin{scope}[shift={(-6,0)}]
\draw(0,0)--(0,2)--(2,2)--(2,0)--(0,0);
\draw(1,1)--(1,3)--(3,3)--(3,1)--(1,1);
\draw(0,0)--(1,1);
\draw(0,2)--(1,3);
\draw(2,2)--(3,3);
\draw(2,0)--(3,1);
\draw[Red,fill=White](0,0)circle(0.2)node{+};
\draw[Red,fill=White](0,2)circle(0.2)node{+};
\draw[Red,fill=White](2,2)circle(0.2)node{+};
\draw[Red,fill=White](2,0)circle(0.2)node{+};
\draw[Blue,fill=White](1,1)circle(0.2)node{-};
\draw[Blue,fill=White](1,3)circle(0.2)node{-};
\draw[Blue,fill=White](3,3)circle(0.2)node{-};
\draw[Blue,fill=White](3,1)circle(0.2)node{-};
\draw(1,-0.75)node{$(A_1,\,A_2,\,A_3^{(1)})$};
\end{scope}
\begin{scope}[shift={(-6.5,1)}]
\draw[-latex](-2,0)--(-2,0.75);
\draw[-latex](-2,0)--(-1.25,0);
\draw[-latex](-2,0)--(-1.5,0.5);
\draw(-2,1)node{$x^2$};
\draw(-1,0)node{$x^1$};
\draw(-1.25,0.75)node{$x^3$};
\end{scope}
\draw(5.5,0)circle(0);
\end{tikzpicture}
\caption{Inequivalent Type II orientifold configurations on $T^3$ with coordinates $(x^1,\,x^2,\,x^3)$ dual to the $Spin(32)/\mathbb{Z}_2$ heterotic string on $T^3$ with the specified Wilson lines (see eqs. \eqref{WL12} and \eqref{WL3}). The $+$ sign corresponds to a stack of 4 $D7$-brane pairs on top of an $O7^-$-plane. The $-$ sign corresponds to a naked $O7^-$-plane. Alternatively, the $+$ signs can be interpreted as $O7^+$-planes, leading to rank reduction by 16 in two inequivalent ways. These correspond to inequivalent ways of freezing four $D_4$ singularities in the M-Theory dual K3. }
\label{fig:or7}
\end{figure}

We will restrict ourselves to the first two configurations in \eqref{WL3}. They lie in the region of the moduli space which can be described perturbatively by Type II orientifolds on $T^3$ (see Figure \ref{fig:or7}), where one can then exchange each $O7^- + 4\, D7$ stack by an $O7^+$ plane, giving two different configurations with rank reduction by 16. These were identified in \cite{deBoer:2001wca} as perturbatively inequivalent, with the condition for full nonperturbative inequivalence being that the corresponding lattice embeddings of $2\,D_4$ in $\Gamma_{3,19}$ be distinct. We have already proven this later statement above, where the inequivalence in the root lattice embeddings is captured by the topology of the corresponding gauge group. Moreover, this second configuration gives rise to a rank 3 moduli space naturally identified with the circle compactification of the rank 2 component in F-Theory $\mathcal{M}_{2D_8}'$, which is perhaps easier to see by using $A_3^{(3)}$ instead of $A_3^{(2)}$ in \eqref{WL3}.
 
Let us refer to the moduli spaces discussed here as $\mathcal{M}_{4D4}$ and $\mathcal{M}_{4D4}'$, respectively. As before, we take the charge lattices to be the orthogonal complements of the weight lattices of the generic gauge groups prior to freezing, obtaining
\begin{eqnarray}
	\begin{split}
		\Gamma_{4D4} &= \Gamma_{1,1}\oplus \Gamma_{2,2}(2)\,,\\
		\Gamma_{4D4}' &= \Gamma_{3,3}(2)\,.\\
	\end{split}
\end{eqnarray}
Both of these are obtained from $\Gamma_{2D_8}$ and $\Gamma_{2 D_8}'$ (cf. Eqs. \eqref{gamma1}, \eqref{gamma2}) by adding a $\Gamma_{1,1}(2)$ factor, as usual. In the first case, there is one maximal symmetry enhancement coming from
\begin{equation}
	\frac{Spin(8)^4\times SU(2)^3}{\mathbb{Z}_2^4}\,,
\end{equation}
with $\mathbb{Z}_2^4$ generated by $\{(s,s,s,s,0,0,0)\,,~(c,c,c,c,0,0,0)\,,~(0,s,c,v,0,1,1)\,,~(0,c,v,s,1,0,1) \}$ \cite{Fraiman:2021soq}, which is of the form
\begin{equation}
	G = \frac{SU(2)^3}{\mathbb{Z}_2^2}\,, ~~~~~ \mathbb{Z}_2^2 = \{(0,0,0),(1,1,0),(0,1,1),(1,0,1)\}\,.
\end{equation}
By giving mass to any of the $SU(2)$ gauge bosons, it breaks to $SO(4)$, in agreement with the results of Section \ref{sss:M81}. Again note that, as in $\mathcal{M}_{2D_8}$, all of the $SU(2)$'s enter symmetrically into $G$. In the second case we have a maximal enhancement 
\begin{equation}
	G = SU(2)^3\,,
\end{equation}
coming from $Spin(10)^3\times Spin(8)/\mathbb{Z}_2$ with $\mathbb{Z}_2 = \{(0,0,0,0),(v,v,v,v)\}$.

\subsection{Components with frozen $3\,E_6$}

The case of $3\, E_6$ differs from $4\, D_4$ insofar as the associated root lattice has rank 18 and so requires fixing not only Wilson line but also metric and B-field moduli in the dual heterotic frame. There are two sublattices of $\Gamma_{3,19}$ with root sublattice of ADE type $3\, E_6$ (cf. Appendix \ref{app:lats}), hence two different moduli space components associated to this freezing.

The first moduli space component corresponds to the weight lattice 
\begin{equation}
 W_{E_6^3/\mathbb{Z}_3}\,,
\end{equation}
with $\mathbb{Z}_3$ diagonal. This lattice actually admits also an embedding into $\Gamma_{2,18}$ \cite{Shimada2000OnEK,Font:2020rsk} so that it can be realized with $A_3 = 0$, and in this sense is analogous to the first $4\, D_4$ moduli space discussed in the previous section. Taking its orthogonal complement in $\Gamma_{3,19}$, we find the charge lattice
\begin{equation}
	\Gamma_{3E_6} = A_2(-1)\oplus \Gamma_{1,1}\,,
\end{equation}
which matches the charge lattice of F-Theory on $(T^4\times S^1)/\mathbb{Z}_3$ \cite{deBoer:2001wca}. Lastly, the gauge group $E_6^3/\mathbb{Z}_3$ in the standard component can only be enhanced to $(E_6^3/\mathbb{Z}_3) \times SU(2)$, and correspondingly, we find only one symmetry enhancement $U(1) \to SU(2)$. In analogy with the moduli space component associated to $E_6^2$, we expect that this $SU(2)$ will be at level 3. It would be good to confirm this explicitly. 

The second component corresponds to 
\begin{equation}
	W_{E_6^3}\,,
\end{equation}
which can be realized by appropriately turning on $A_3$ in the heterotic frame. The charge lattice is found to be
\begin{equation}
	\Gamma_{3E_6}' = A_2(-1) \oplus \Gamma_{1,1}(3)\,,
\end{equation}
which has a symmetry enhancement $U(1) \to SU(2)$ associated to the standard component gauge group $E_6^2 \times E_7$. Perhaps this theory is dual to F-Theory on $(T^4 \times S^1)/\mathbb{Z}_3$ with a discrete modulus field turned on in such a way that is compatible with the orbifold symmetry.

\subsection{Components with frozen $D_4+2\,E_7$}

The case of $D_4 + 2\, E_7$ is similar to that of $3\, E_6$, having two different moduli space components. The first is associated to the lattice
\begin{equation}
	W_{Spin(8)\times E_7^2/\mathbb{Z}_2}\,,
\end{equation}
which can be realized with $A_3 = 0$. The charge lattice is
\begin{equation}
	\Gamma_{D_4+2E_7} = 2\, A_1(-1) \oplus \Gamma_{1,1}\,,
\end{equation}
matching that of F-Theory on $(T^4 \times S^1)/\mathbb{Z}_4$. There are two symmetry enhancements of the type $U(1) \to SU(2)$, coming respectively from the standard component gauge groups $(Spin(8)\times E_7^2)/\mathbb{Z}_2 \times SU(2)$ and $(Spin(10) \times E_7^2) / \mathbb{Z}_2$ with $\mathbb{Z}_2$ diagonal \cite{Fraiman:2021soq}.

The second component is associated to the lattice
\begin{equation}
 W_{Spin(8)\times E_7^2}\,,
\end{equation}
and has charge lattice
\begin{equation}
	\Gamma_{D_4+2E_7}' = 2\, A_1 (-1) \oplus \Gamma_{1,1}(2)\,.
\end{equation}
It also has two symmetry enhancements, which seem to be $U(1) \to SU(2)$ and $U(1) \to SO(3)$, coming respectively from $Spin(10)\times E_7^2$ and $Spin(8)\times E_7 \times E_8$. An embedding of the simple root for $SO(3)$ into $\Gamma_2$ can be taken as
\begin{equation}
	\ket{\alpha} = \ket{i\sqrt{2},0;1,2}\,,
\end{equation}
where the first two components realize an $A_1(-1)$ each, using the inner product
\begin{equation}
	\braket{a,b;n,m|a',b';n,m'} = aa'+bb'+nm'+n'm\,.
\end{equation} 
It is easy to see that there are no vectors in $\Gamma_2$ with inner product 1 with $\ket{\alpha}$, hence it corresponds to $SO(3)$ and not $SU(2)$. This conclusion hinges on interpreting all norm 2 vectors as roots corresponding to massless states, which depending on finer details of the theory may not be the case. 

\section{Uplift to 9d}
\label{sec:4}
When theories with rank reduction are further compactified on a circle, their charge lattices get extended by a lattice of the form $\Gamma_{1,1}(n)$. From the results of \cite{deBoer:2001wca}, we have that $n = 2,3,4,6$ for the 7d theories related to freezing the singularities $4\, D_4$, $3\, E_6$, $D_4+ 2 \,E_7$ and $D_4+E_6+E_8$, respectively. The alternative component for the first of these singularities that we have proposed has charge lattice $\Gamma_{3,3}(2)$, which uplifts to $\Gamma_{2,2}(2)$ upon decompactification to $\mathcal{M}_{2D_8}'$. The logic here outlined clearly suggests that this procedure can be carried upwards to a nine dimensional theory with charge lattice $\Gamma_{1,1}(2)$.

To add more weight to this conjecture, note that the mechanism of singularity freezing is sensible only to the gauge algebra in eight and seven dimensions, and that the frozen singularity of type $4\,D_4$ in 7d uplifts to $2\, D_8$ in 8d. This uplifting is related to the fact that the later has a Dynkin diagram which is embedded into the affine Dynkin diagram of the former \cite{Witten:1997bs}. Then, $2\, D_8$ can be further uplifted to either $2\, E_8$ or $D_{16}$\footnote{The results of \cite{Lee:2021usk} indicate that at long distance limits in the moduli space of elliptic K3 complex structures only affine groups of type $\hat E_n$ appear, but $\hat D_{16}$ does seem to appear very naturally from the point of view of the dual heterotic string \cite{Collazuol:2022jiy}.}, and that there are two possibilities matches nicely with the fact that $\mathcal{M}_{2D_8}$ has two rank 1 decompactification limits (see \cite{Cvetic:2022uuu} for a verification from the point of view of string junctions). What we argue here is that as well as for $2\, D_8$, there are two inequivalent theories associated to $D_{16}$, the difference being manifest in the generic gauge group associated to the configuration of the given theory prior to reducing the rank.

No singularity freezing mechanism is known in the case of nine dimensional theories, but it is not hard to propose one\footnote{As in the case of F-Theory with frozen singularities on an elliptic K3, this procedure is formal. An explicit understanding of what it means to freeze a singularity in these theories is not available at this point.}. In \cite{Cachazo:2000ey} a nonperturbative description of the Type I' was proposed to be given by the geometry of a real elliptic K3 surface, emerging naturally as a decompactification limit of an elliptic K3 in a similar way to how an elliptic K3 emerges from a generic K3 surface. On it, the gauge groups $Spin(32)/\mathbb{Z}_2$ and $Spin(32)$ can be realized, and indeed a motivation for this construction was the fact that these gauge groups are not distinguished at the perturbative level from the Type I' picture. We can then define a frozen singularity as an ADE singularity in the real K3 of type $E_8$ or $D_{16}$, which would correspond in the Type I' picture respectively to a neutral $O8$-plane \cite{Aharony:2007du} or an $O8^+$-plane. In the later case, a situation similar to the one in eight dimensions would hold, producing two inequivalent rank 1 moduli spaces with charge lattices $\Gamma_{1,1}$ and $\Gamma_{1,1}(2)$ respectively.

This reasoning implies that in nine dimensions there are three inequivalent nonabelian gauge groups of rank 1, each one corresponding to a different theory. In fact, this matches nicely with the results in Table 4 of \cite{Bedroya:2021fbu}, whereby Swampland considerations lead to three possible such gauge groups, denoted by $A_1$, $E_1$ and $C_1$. We argue that the later two are naturally identified with the frozen singularities $Spin(32)/\mathbb{Z}_2$ and $Spin(32)$, respectively, and are not connected one to another. This expectation matches an analysis of potential discrete theta angles in nine dimensional string compactifications with $\mathcal{N} = 1$ \cite{MP}. Our results imply the existence of new string compactifications which do exist and are the subject of the companion paper \cite{MP}.

\section{Conclusions and outlook}
In this note we have shown that if the mechanism of singularity freezing in F-Theory and M-Theory on K3 surfaces is always allowed whenever the appropriate ADE singularities are present, these lead to disconnected moduli spaces in certain cases. We have furthermore argued that this picture should extend to nine dimensions using real elliptic K3 surfaces which were proposed to give a nonperturbative description of the physics partially captured by Type I' string theory. 

There is still the necessity of having a more rigorous understanding of how ADE singularities can be frozen in different (geometrical descriptions of) string theories. It is likely that one could formulate a framework in which the Narain lattices play a main role, as they seem to encode a great amount of information regarding the overall structure of the moduli space including all connected components (of course, in the case of 16 supercharges). We note however that such a framework should include more information that, in particular, specifies the ways in which the 3-form field can be turned on in M-Theory on a K3 surfaces; there are two components in seven dimensions associated to the lattice $2\, E_8$, distinguished by this choice of 3-form field for the background.

On the other hand, classifying the moduli space components of 6D $\mathcal{N} = (1,1)$ vacua is still an open problem for which conventional tools have are difficult to apply \cite{deBoer:2001wca}. A potential solution based also on techniques involving charge lattices will be reported in \cite{FP}.

\subsection*{Acknowledgements}
We are grateful to Peng Cheng, Veronica Collazuol, Jacob McNamara, Cumrun Vafa and specially Bernardo Fraiman and Miguel Montero for helpful discussions. We thank Mariana Graña for useful discussions and comments on the manuscript. This work was supported by the ERC Consolidator Grant 772408-Stringlandscape.

\appendix
\section{Uniqueness of lattice embeddings}
\label{app:lats}

In this appendix we consider the problem of uniqueness of a primitive embedding of a lattice $M$ into either Narain lattices which defines a connected component in moduli space. To this end we make use of the discriminant group $A_M$ of $M$, defined as the quotient $M^*/M$, and the discriminant quadratic form $q_M$, which is defined as the extension of the quadratic form on $M$ to $M^*$ mod 2 and assigns a rational number mod 2 to each element $x$ of $A_M$. A powerful theorem of Nikulin \cite{NikulinVV1980ISBF} then allows to immediately solve our problem for most of these lattices:\\

\noindent \textbf{Nikulin's Theorem:} Let $L$ be an even unimodular lattice with signature $(s_+, s_-)$, $s_+ > 0$, $s_- > 0$, and $L$ an even lattice with signature $(t_+,t_-)$ and discriminant quadratic form $q_M$. Let $l(A_M)$ be the minimun number of generators of $A_M$. Let the following inequalities be satisfied:
\begin{equation}
	\begin{split}
	t_+ &< s_+\,, \\
	t_- &< s_-\,, \\
	\text{rank}(M) + l(A_M) &\leq \text{rank}(L)-2\,.
	\end{split}
\end{equation}
Then there exists an unique primitive embedding of $M$ into $L$.\\

In Table \ref{tab:nik} we list the lattices $M$ together with the data related to Nikulin's theorem. Note that these lattices are just the weight lattices of the gauge groups listed in Table \ref{tab:frozen}. As we can see, the theorem fails only for the lattices related to the new moduli spaces proposed here. The reason is simply that these lattices correspond to gauge groups with simpler fundamental groups and thus have larger values for $\ell(M)$. It can be proven by hand that the embedding of $2\, D_8$ into $\Gamma_{2,18}$ is indeed unique by exploiting the fact that $D_8+D_7$ does satisfy Nikulin's theorem and checking for a particular representative of this lattice all the extensions to $2\, D_8$ are equivalent. For the other components, this analysis is much more involved and inconclusive. As explained in the text for $4\, D_4$, however, inequivalent embeddings of the exact same weight lattice into the Narain lattices should produce the same physical theory. Nonetheless, having control on the uniqueness of these embeddings would still be desirable.  

\begin{table}
	\centering
\def\arraystretch{1.3}
\begin{tabular}{|c|c|c|c|}\hline
$M$ & $A_M$ &  $\text{rank}(M) + l(A_M)$ & Theorem applies for $\Gamma_{2,18}$\\ \hline 
$D_{8}$&$\mathbb{Z}_{2}^2$ & $10$ & \cmark\\ \hline
$[2\,D_{8}]^+$&$\mathbb{Z}_2^2$ & $18$ & \cmark\\ \hline
$2\,D_{8}$&$\mathbb{Z}_2^4$ & $20$ & \xmark\\ \hline\hline
$M$ & $A_M$ &  $\text{rank}(M) + l(A_M)$ & Theorem applies for $\Gamma_{3,19}$\\ \hline 
$2\,D_4$&$\mathbb{Z}_2^4$ & $12$ & \cmark\\ \hline
$2\,E_6$&$\mathbb{Z}_3^2$ & 14& \cmark\\ \hline
$2\,E_7$&$\mathbb{Z}_2^2$ & 16 & \cmark\\ \hline
$2\,E_8$&$\mathbb{Z}_1$ & 17  & \cmark\\ \hline
$[4\, D_4]^{++}$&$\mathbb{Z}_2^4$ & $20$ & \cmark\\ \hline
$[4\, D_4]^{+}$&$\mathbb{Z}_2^6$ & $22$ & \xmark\\ \hline
$[4\, D_4]^{+'}$&$\mathbb{Z}_2^6$ & $22$ & \xmark\\ \hline
$[3\,E_6]^+$&$\mathbb{Z}_3$ & 19 &\cmark\\ \hline
$3\,E_6$&$\mathbb{Z}_3^3$ & 21 & \xmark\\ \hline
$[D_4+2\, E_7]^+$&$\mathbb{Z}_2^2$ & 20 & \cmark\\ \hline
$D_4+2\, E_7$&$\mathbb{Z}_2^4$ & 22 & \xmark\\ \hline
$D_4+ E_6 + E_8$&$\mathbb{Z}_2\times \mathbb{Z}_3$ & 20 & \cmark\\ \hline
\end{tabular}
\caption{Lattices which define moduli space components of string theories with 16 supercharges in eight and seven dimensions, embedded respectively into $\Gamma_{2,18}$ and $\Gamma_{3,19}$, together with their discriminant groups and values for the sum of the rank and minimal number of discriminant group generators. The last column indicates if the embedding uniqueness is guaranteed by Nikulin's theorem. The pluses indicate extensions to weight lattices (cf. Table \ref{tab:frozen}.)}\label{tab:nik}
\end{table}

\newpage
\bibliographystyle{JHEP}
\bibliography{newstrings}

\end{document}